# Density-Functional Theory Study of the Optoelectronic Properties of π-Conjugated Copolymers for Organic Light-Emitting Diodes

Tarek Mestiri*
Unité de Recherche: Matériaux Nouveaux et Dispositifs Électroniques Organiques UR 11ES55,
Faculté des Sciences de Monastir, Université de Monastir, Monastir, Tunisia

Ala Aldin M. H. M. Darghouth†
Laboratoire de Chimie Théorique, Département de Chimie Moléculaire (DCM),
Institut de Chimie Moléculaire de Grenoble (ICMG),
Université Grenoble-Alpes, 301 rue de la Chimie, CS
40700, 38058 Grenoble Cedex 9, France

Mark E. Casida‡
Laboratoire de Chimie Théorique, Département de Chimie Moléculaire (DCM),
Institut de Chimie Moléculaire de Grenoble (ICMG),
Université Grenoble-Alpes, 301 rue de la Chimie, CS
40700, 38058 Grenoble Cedex 9, France

Kamel Alimi§
Unité de Recherche: Matériaux Nouveaux et Dispositifs Électroniques Organiques UR 11ES55,
Faculté des Sciences de Monastir, Université de Monastir, Monastir, Tunisia

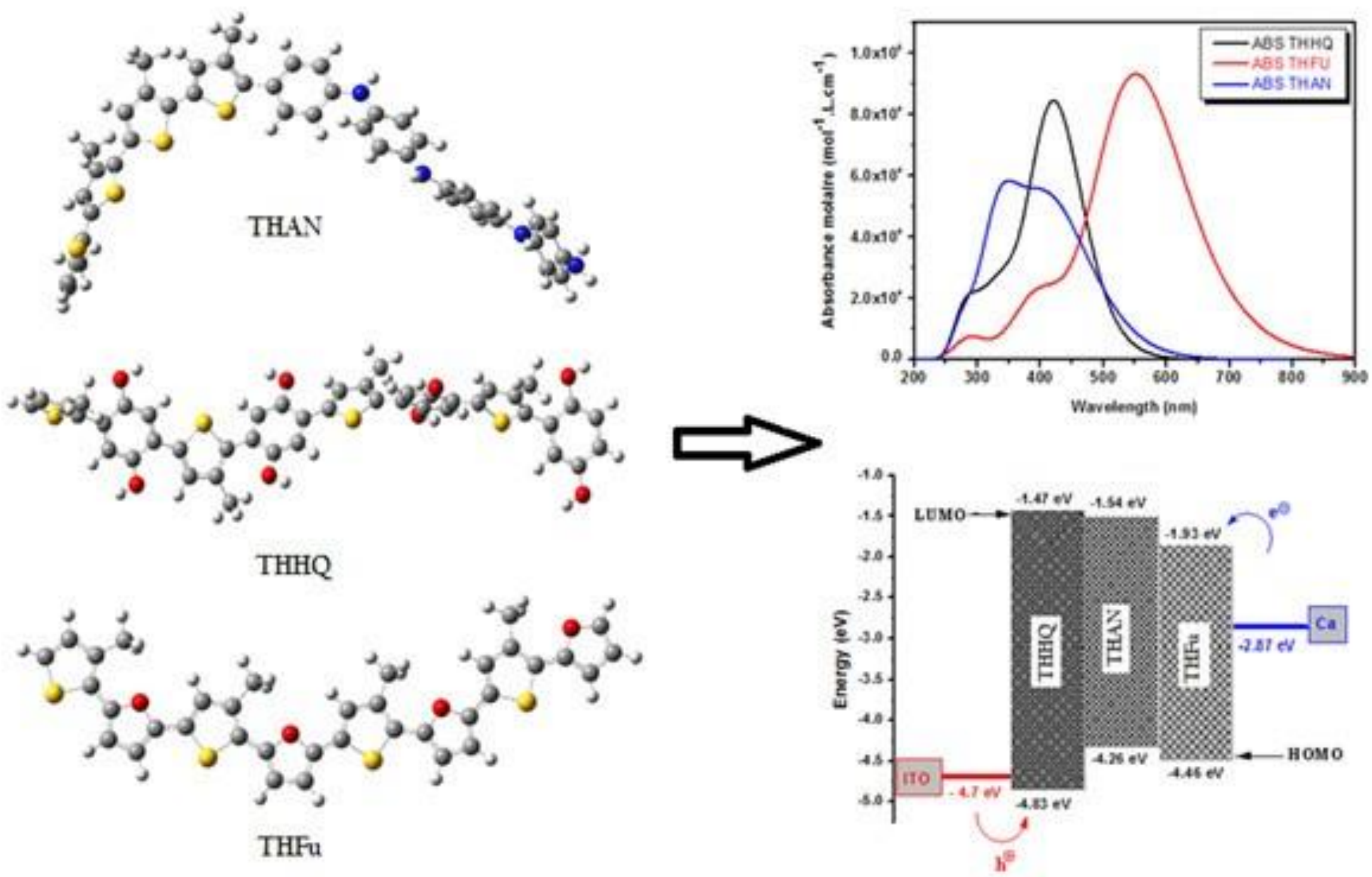

Novel low-band-gap copolymer oligomers are proposed on the basis of density functional theory (DFT) quantum chemical calculations of photophysical properties. These molecules have an electron donor-accepter (D-A) architecture involving poly(3-hexylthiophene-2,5-diyl) (P3HT) as D units and furan, aniline, or hydroquinone as A units. Structural parameters, electronic properties, highest occupied molecular orbital (HOMO)-lowest unoccupied molecular orbital (LUMO) gaps and molecular orbital densities are predicted. The charge transfer process between the D unit and the A unit one is supported by analyzing the optical absorption spectra of the compounds and the localization of the HOMO and LUMO.

## I. INTRODUCTION

Among the many marvels of modern life is the increasing integration of organic components into modern electronics. This revolution is taking place because organic electronics can be relatively inexpensive to make and to cast, can be printable, and may profit from a large body of organic chemistry to make new and interesting electronic materials[1–4]. One of the currently most popu-lar example of organic electronics is the organic light emitting diode (OLED)[5,6] which is shown schematically in Fig. 1. The goal of this letter is to report a theoretical study of some novel candidate materials made of oligomers $(DA)_n$ and $D_nA_n$ of co-polymers composed of electron donor (D) and electron acceptor (A) units. It may be regarded as pre-screening work prior to experimental synthesis and testing of the most likely candidates. In our case the D unit is always poly(3-hexylthiophene-2,5-diyl [P3HT, also known as poly(3-hexylthiophene)] while the acceptor unit is varied between furane, aniline, and hydroquinone (see Fig. 2).

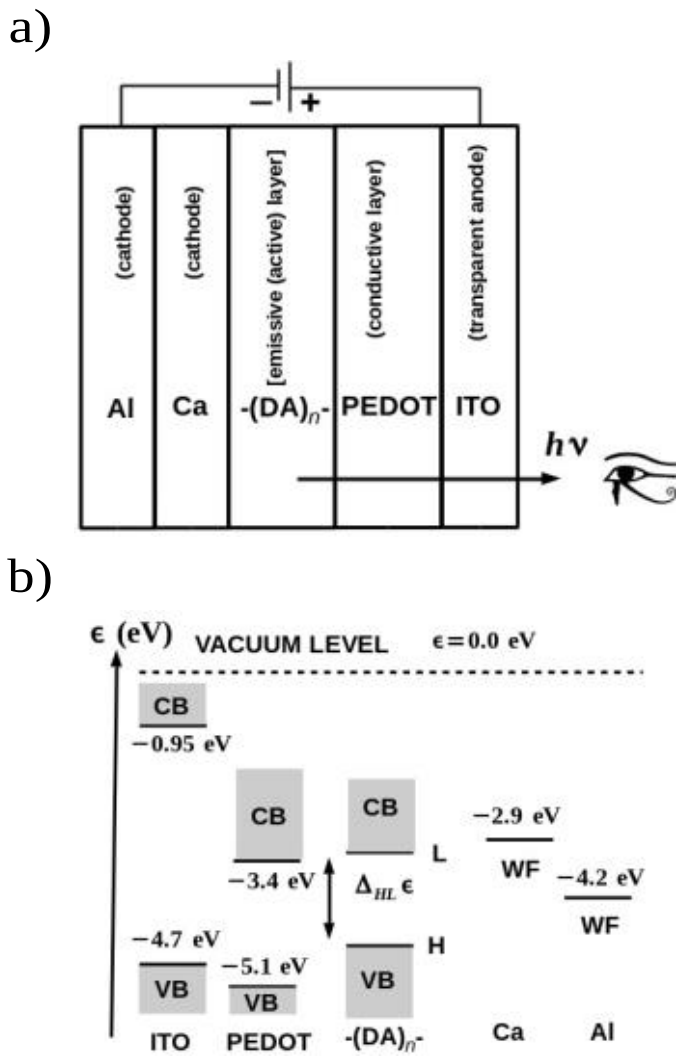

FIG. 1: A typical OLED: (a) Electrons are injected by the anode (here represented by aluminum and calcium) while holes are injected by a transparent cathode (ITO, indium tin oxide) into a conductive layer (here represented by PEDOT, polyethylenedioxythiophene). The electrons and holes combine in an active layer (here represented by a polymer or oligomer of alternating donor D and acceptor A units) where photons are created and emitted. (b) Band diagram for the different components in the OLED. The anode work functions (WFs) are indicated on the right. The anode and conductive layer valence band (VBs) and conductive bands (CBs) are indicated on the right. They are also indicated for the active layer where the difference between the highest-occupied molecular orbital (HOMO or just H) and lowest-unoccupied molecular orbital (LUMO or just L) is the HOMO-LUMO gap, $\Delta_{HL} = {}_{L} - {}_{H}$.

There are several criteria which need to be met by the active layer. Let us list a few: The active layer should be a semiconductor, which means that its band gap should be lower than that for a typical insulator but it should not be a conductor (i.e., with zero band gap). The lowest-unoccupied molecular orbital (LUMO or just L) energy should be lower than the typical anode work function ($<$ ca. -3 eV) and the highest-occupied molecular orbital (HOMO or just H) energy should be higher than the valence band (VB) of the conducting layer ($>$ ca. -5 eV for PEDOT, polyethylene-dioxythiophene). We would also like emission to occur in the visible (Fig. 3). The values of the anode work function and of the valence band of the conducting layer suggest emission at around 2 eV which is in the red, but is modified by adding an applied voltage. Emission in the blue is more useful than emission in the red because it may be used to excite still other layers which emit at lower wave lengths. This is possible with even a small HOMO-LUMO gap as long as the HOMO is taken as the upper limit of the VB while the LUMO is the lower limit of the conduction band (CB). Also, it is important to keep in mind that the schema shown in Fig. 1 is only approximate. Strictly speaking, band energies correspond to electron removal and electron addition energies, so the number of electrons is changing. Optical gaps correspond to an excitation at fixed number of electrons. Furthermore, we shall use density-functional theory (DFT) to determine orbital energies which, strictly speaking, correspond neither to electron removal, electron addition, or optical excitation energies. However they provide a useful first approximation.

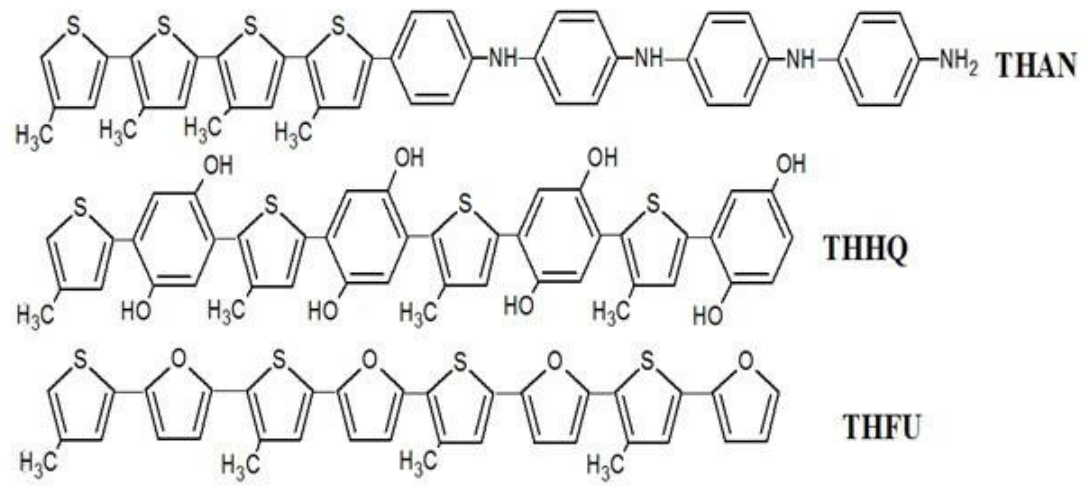

FIG. 2: Molecules considered in this paper: an oligomer of a block polymer composed of a tetramer of P3HT and of a tetramer of aniline (THAN); an oligomer of an alternating copolymer of P3HT and furane (THHQ); and an oligomer of an alternating copolymer of P3HT and hydrogquinone (THHQ).

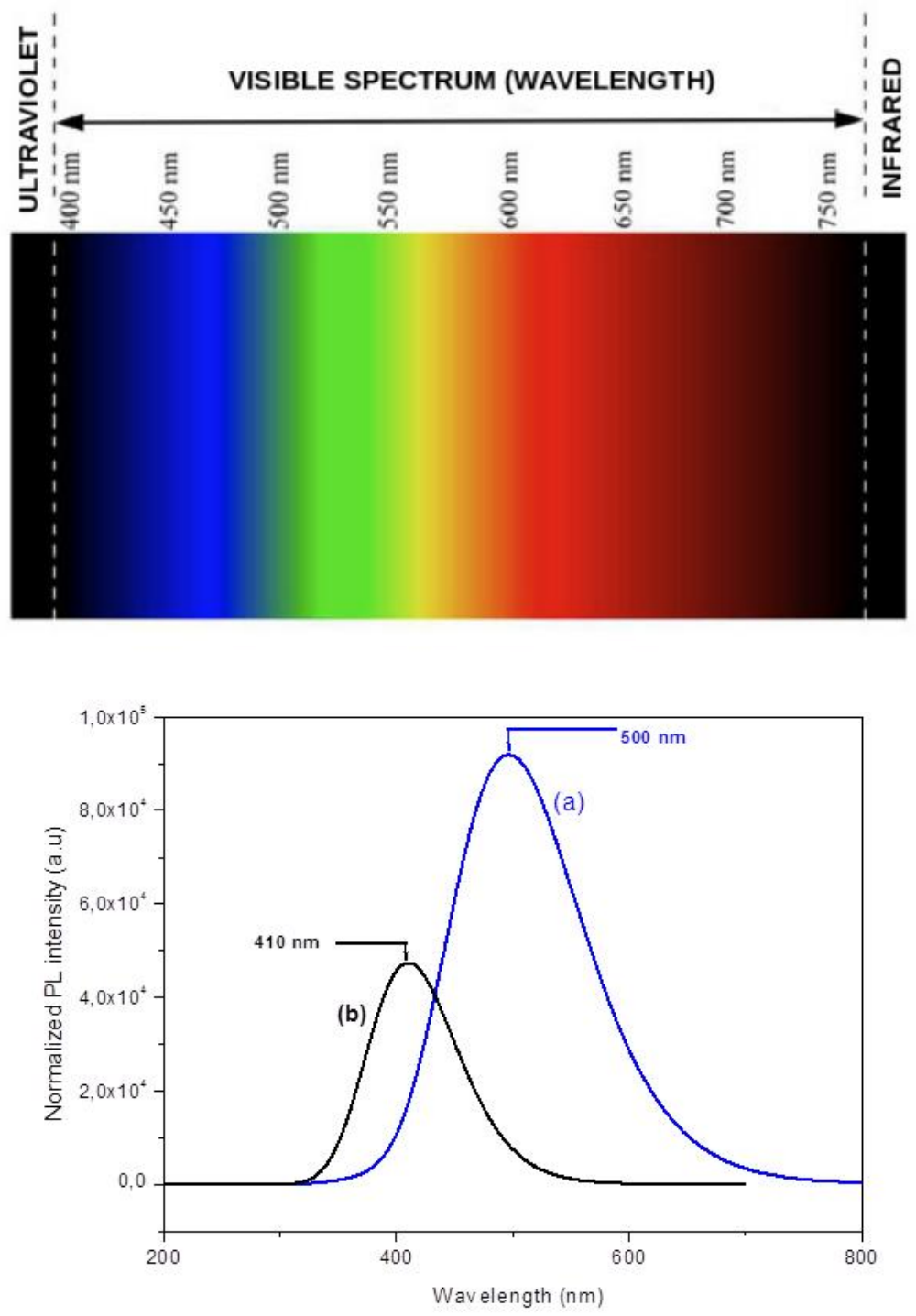

FIG. 3: Theoretical PL spectra of copolymer THHQ (a) compared with the oligomer one PHQ (b). Visible spectrum [based upon a figure from the Center of Higher Education and Initation to Experimental Research of the University Joseph Fourier (Grenoble)]: red 617–751 nm (1.65–2.01 eV); orange 588–617 nm (2.01–2.11 eV); yellow 571–588 nm (2.11–2.17 eV); green 496–571 nm (2.17–2.50 eV); blue 451–496 nm (2.50–2.75 eV); violet 379–451 nm(2.75–3.27 eV).

The choice of molecules (Fig. 2) whose calculated properties are reported in this paper is governed by $FeCl_3$ oxidative coupling reactions as has already been extensively used in Monastir[7–9]. As these molecules have yet to be synthesized, we cannot as yet say how easy they will be to synthesize. Further details may be found in the Supplementary Material associated with this article.

The rest of this letter is organized as follows: The next section provides computational details. This is followed by a results section where we give the principal results (other results being relegated to the Supplementary Information associated with this letter). We finish with a concluding discussion.

## II. COMPUTATIONAL DETAILS

All electronic structure calculations were carried out using the Gaussian09 program[10].

Monomer geometric structures are fully optimized using the popular B3LYP functional. This functional has the form,

$$\begin{aligned} E_{xc} = & (1-a_0)E_x^{LDA} + a_0 E_x^{HF} + a_x E_x^{B88x} \\ & + a_c E_c^{LYP88c} + (1-a_c)E_c^{LDA}, \end{aligned} \quad (2.1)$$

where $a_0 = 0.20$, $a_x = 0.72$, and $a_c = 0.81$ are taken from the B3P functional[11], LDA stands for the local density approximation[12] with the Vosko-Wilk-Nusair parameterization[13], B88x is Becke's 1988 generalized gradient approximation (GGA) for exchange[14], and LYP88c is Lee, Yang, and Parr's 1988 GGA for correlation[15]. Calculations were either performed using the 6-31G split valence basis set[16] (i.e., using the B3LYP/6-31G model chemistry) or with the same basis set augmented with d-type polarization functions on the heavy atoms[17] (i.e., using the B3LYP/6-31G(d) model chemistry). Dimer structures were optimized by carrying out a scan of the torsional energy ciurves of each molecule at the B3LYP/6-31G level[18]. Higher oligomers were obtained without further optimzation by joining optimized monomers with the dimer optimized torsional angles.

Electronic properties such as HOMO and LUMO energies, as well as their associated gaps, were also calculated for studied copolymers. Therefore, the optimized geometries were calculated for cationic and anionic charged states. Optical absorption spectra were calculated using the time-dependent density-functional theory at the TD-B3LYP/6-31G level and were fitted to Gaussian curves within Swizard program[19,20]. Electronic transition assignments and oscillator strengths were also calculated using the same method of calculation.

TABLE I: Optimized $M_1$-$M_2$ dimer torsional angles between the rings in monomer $M_1$ and in monomer $M_2$. Optimized torsional angles are good to $\pm 10^\circ$ or better.

| $M_1$-$M_2$ | $M_1$ | $M_2$ | angle |
|---|---|---|---|
| THAN | P3HT | aniline | $100^\circ$ ($60^\circ$ [a]) |
| P3HT | P3HT | P3HT | $60^\circ$ |
| PHQ | P3HT | hydroquinone | $120^\circ$ |
| PFu | P3HT | furane | $140^\circ$($60^\circ$[a]) |

[a] Secondary (local) minimum.

Photoluminescence spectra were calculated by first optimizing the lowest singlet state with the TD-B3LYP/6-31G(d) model chemistry[21,22]. This geometry was then used to calculate transition energies also at the TD-B3LYP/6-31G(d) level.

The simulation of vibrational spectra was performed using the B3LYP/6-31G model.

## III. RESULTS

### A. Geometries

a. Dimers All the dimer structures (oligomer of aniline PANI, copolymer P3HT-aniline THAN, oligomer of hydroquinone PHQ, copolymer P3HT-hydroquinone THHQ, oligomer of furane PFu, and copolymer P3HT-furane THFu) were optimized in the ground state using the B3LYP/6-31G(d) model chemistry (Fig 5). This was done by scanning the dihedral angle between the monomers and simultaneously relaxing all other degrees of freedom. The results are shown in Fig. 4 and in Table I. All dimer geometry optimizations were checked by calculating vibrational frequencies. The calculated infrared spectra are provided in the Supplementary Material.

b. Higher Oligomers The fully optimized structures of the higher oligomers with DFT/B3LYP/6-31G(d) method, with respect to the inter-ring dihedral angles of all the studied structures are shown in Fig. 5.

Our strategy begins with the model structure of the monomer, first, we optimised it and then we calculated the energy gap for the optimised monomer. Second, we increased the number of the units of the model by increasing one unit at each step with optimization. Then, we calculated the energy gap at each step to see how the energy gap affected by the number of the units of the model.

a)

b)

c)

d)

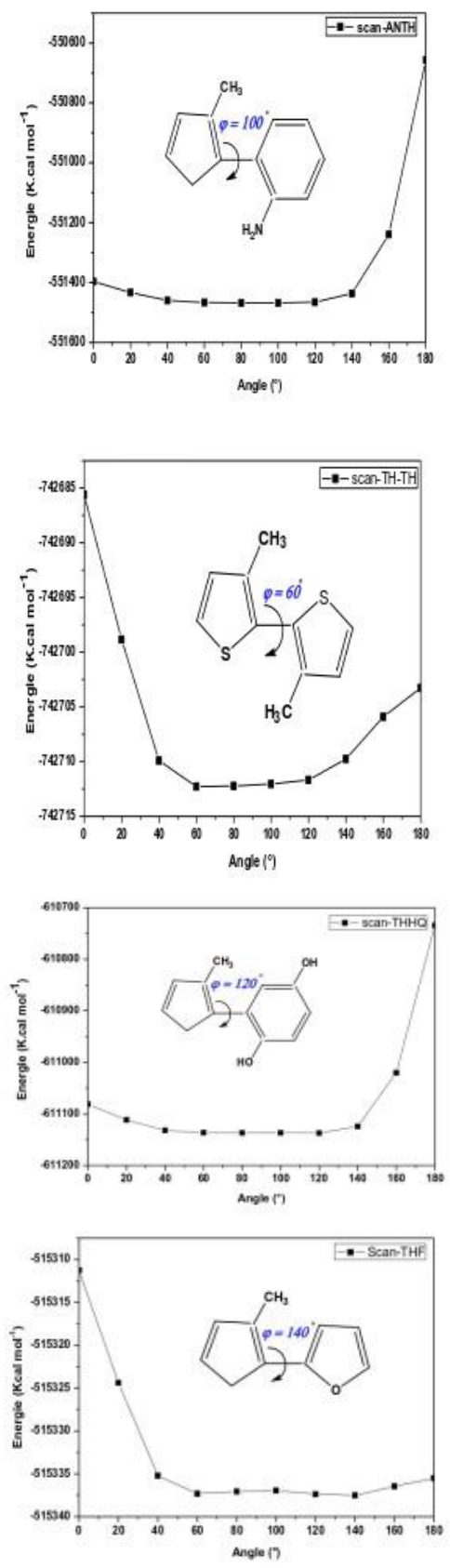

FIG. 4: B3LYP/6-31G(d) torsional energy curves: (a) P3HT-co-P3HT (P3HT), (b) P3HT-co-aniline (THAN), (c) P3HT-co-hydroquinone (THHQ), and (c) P3HT-co-furane (THFu).

So, we found that the energy gap keep decreasing with increasing the number of the units until reach to the constant value as shown in the Fig. 6.

### B. Optical and electronic properties

We are interested in whether integrating a donor unit P3HT into a polymer made up of acceptor units will decrease the band and optical gaps and by how much. In particular, we wish to know: What happens when PANI (the oligomer of aniline) becomes THAN (the co-oligomer of P3HT and of aniline)? What happens when PFu (the oligomer of furane) becomes THFu (the co-oligomer of P3HT and furane)? What happens when PHQ (the oligomer of hydroquinone) becomes THHQ (the co-oligomer of P3HT and hydroquinone)?

$$\begin{aligned} \text{PANI} &\rightarrow \text{THAN} \\ \text{PFu} &\rightarrow \text{THFu} \\ \text{PHQ} &\rightarrow \text{THHQ}\,. \end{aligned} \tag{3.1}$$

Figure 7 shows the density-of-states (DOS) for the various oligomers. Table II shows the HOMO, LUMO, and HOMO/LUMO gaps for the various

a)

b)

c)

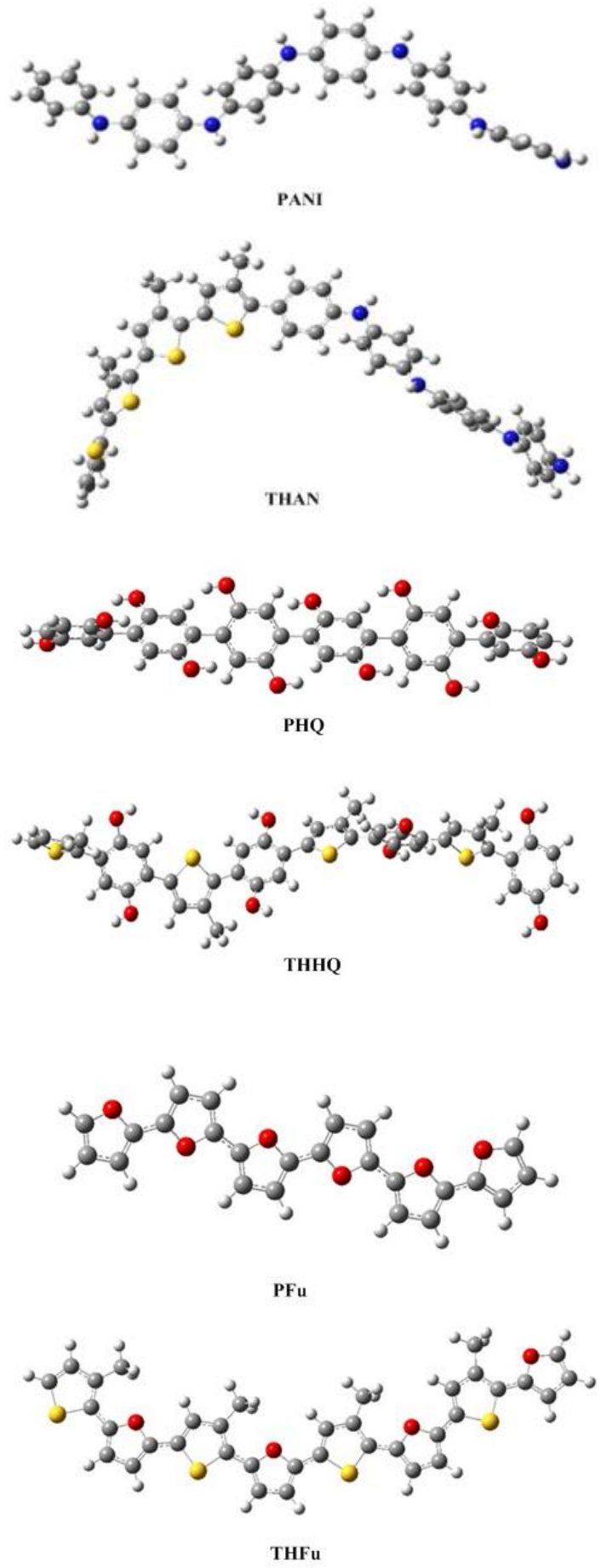

FIG. 5: B3LYP/6-31G(d) optimized geometric structures: (a) PANI and THAN, (b) PHQ and THHQ, and (c) PFu and THFu.

TABLE II: HOMO, LUMO, and gap values for the oligomers and corresponding co-oligomers.

| | HOMO (eV) | LUMO (eV) | $E_g$ (eV) |
|---|---|---|---|
| P3HT | -4.8 | -2.86 | 1.94 |
| | Pani → THAN | | |
| Pani | -4.08 | -0.14 | 3.94 |
| THAN | -4.26 | -1.54 | 2.71 |
| | PHQ → THHQ | | |
| PHQ | -5.56 | -1.55 | 4.01 |
| THHQ | -4.83 | -1.47 | 3.36 |
| | PFU → THFu | | |
| PFu | -4.53 | -1.54 | 2.98 |
| THFu | -4.46 | -1.93 | 2.53 |

oligomers. We see that alternating donor and acceptor units leads to lowering the HOMO/LUMO gap $E_g$ of the oligomer. Intuitively this makes sense because the donor has a high HOMO while the acceptor has a low LUMO. Classic frontier molecular orbital diagram arguments then show that the $E_g$ in the bound D-A system should be smaller than in either the D or A alone (Fig. 8).

Differences of orbital energies are only first approximations to actual absorption energies. Calculated absorption spectra are shown in Fig. 9. The analysis of the TD-B3LYP coefficients allows us to extract the optical gaps (Table III). Figure 10 shows that the optical gap correlates remarkably well with the HOMO/LUMO gap so that reductions in the optical gap may be explained, at least to lowest order, the the same simple same way as reductions in the HOMO/LUMO gap.

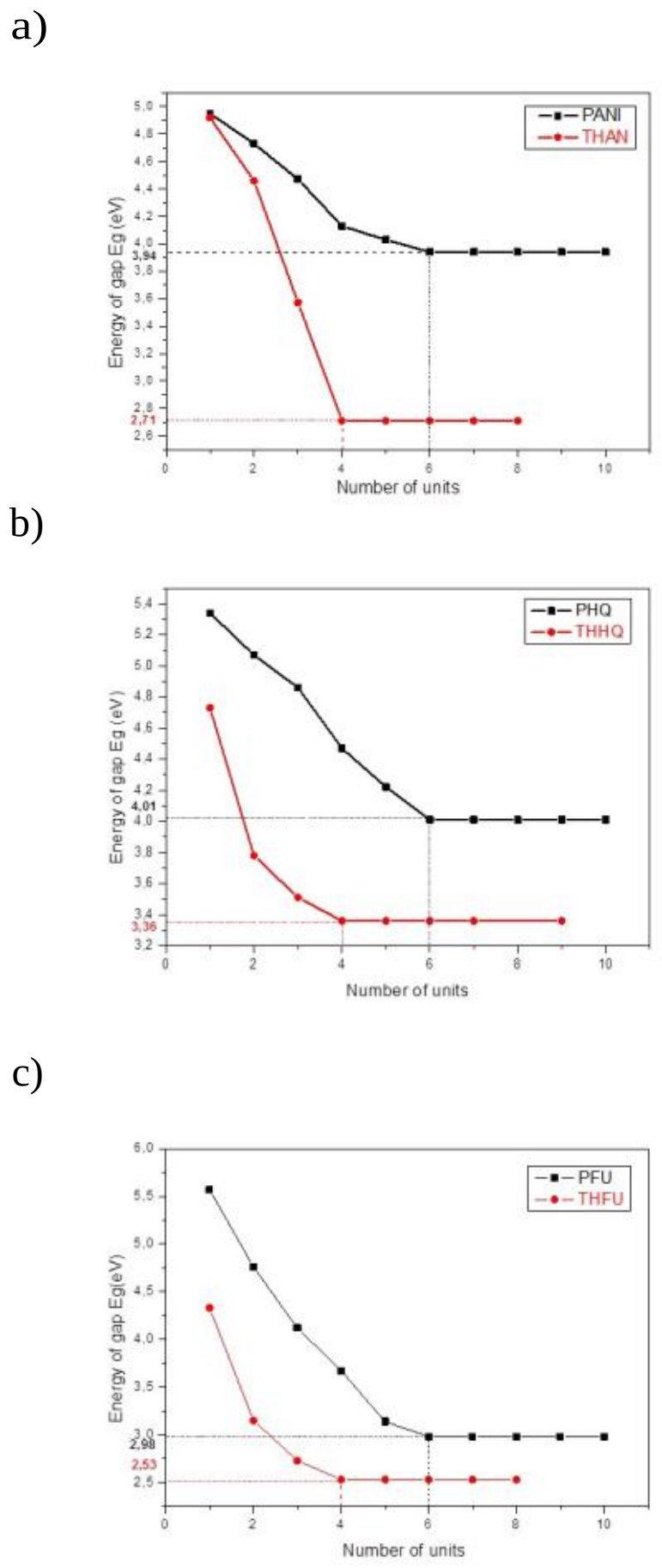

FIG. 6: Evolution of the B3LYP/6-31G(d) HOMO/LUMO gap as a function of the number of monomers: (a) PANI and THAN, (b) PHQ and THHQ, and (c) PFu and THFu.

TABLE III: Optical gaps obtained from TD-B3LYP/6-31G(d).

| Molecule | Wavelength (nm) | Energy (eV) | Oscillator strength (f) | Assignments | |
|---|---|---|---|---|---|
| | | PANI → THAN | | | |
| PANI | 359 | 3.45 | 0.806 | H → L | 90% |
| | | | | H-1 → L | 2% |
| THAN | 421 | 2.94 | 1.679 | H-1 → L | 90% |
| | | | | H-2 → L | 3% |
| | | | | H → L | 5% |
| | | PHQ → THHQ | | | |
| PHQ | 355 | 3.49 | 0.898 | H → L | 88% |
| | | | | H-2 → L | 2% |
| | | | | H-1 → L+1 | 5% |
| THHQ | 423 | 2.93 | 2.032 | H → L | 95% |
| | | | | H-1 → L+1 | 4% |
| | | PFu → THFu | | | |
| PFu | 454 | 2.73 | 1.862 | H → L | 100% |
| THFu | 552 | 2.24 | 2.281 | H → L | 100% |

Interestingly, reducing the optical gap also cor-relates with improved absorbance as shown in Fig. 11.

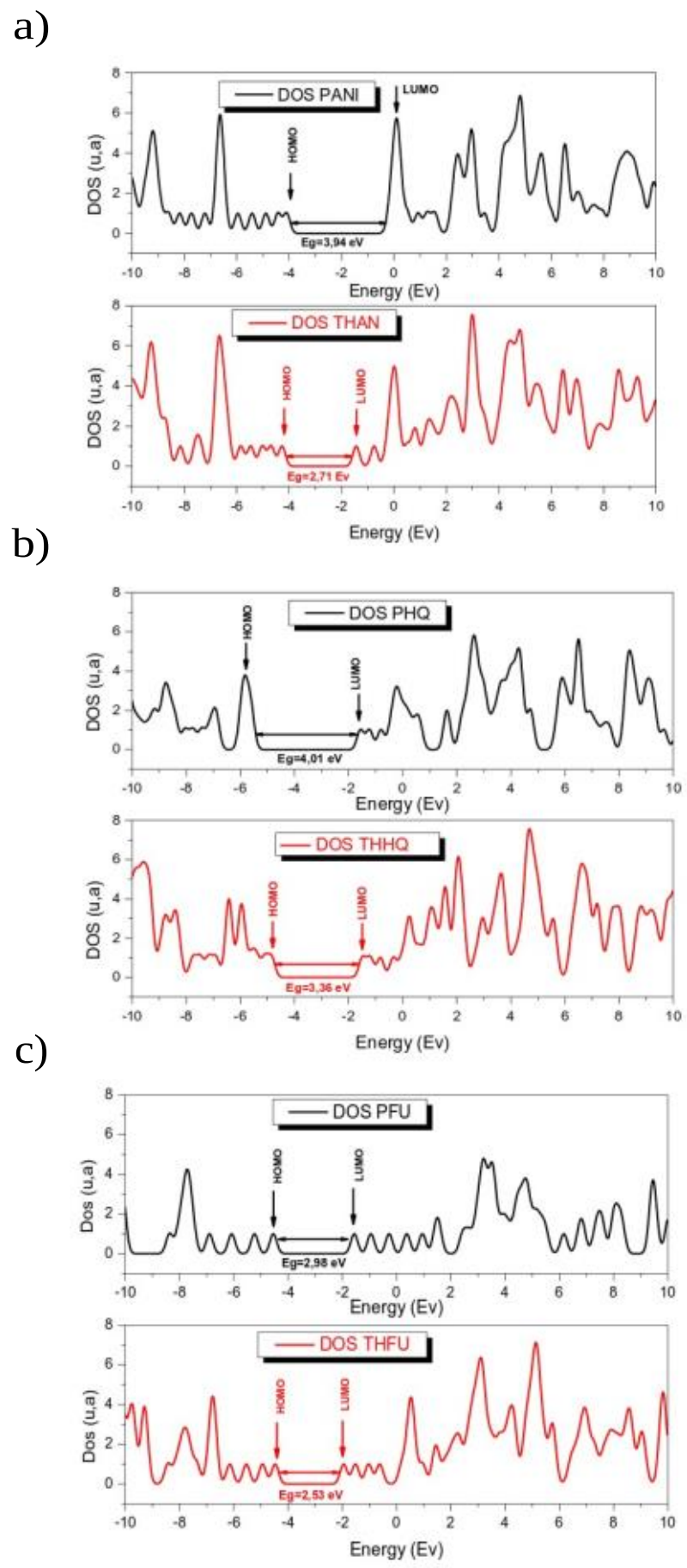

FIG. 7: Density of states: (a) PANI and THAN, (b) PHQ and THHQ, and (c) PFu and THFu.

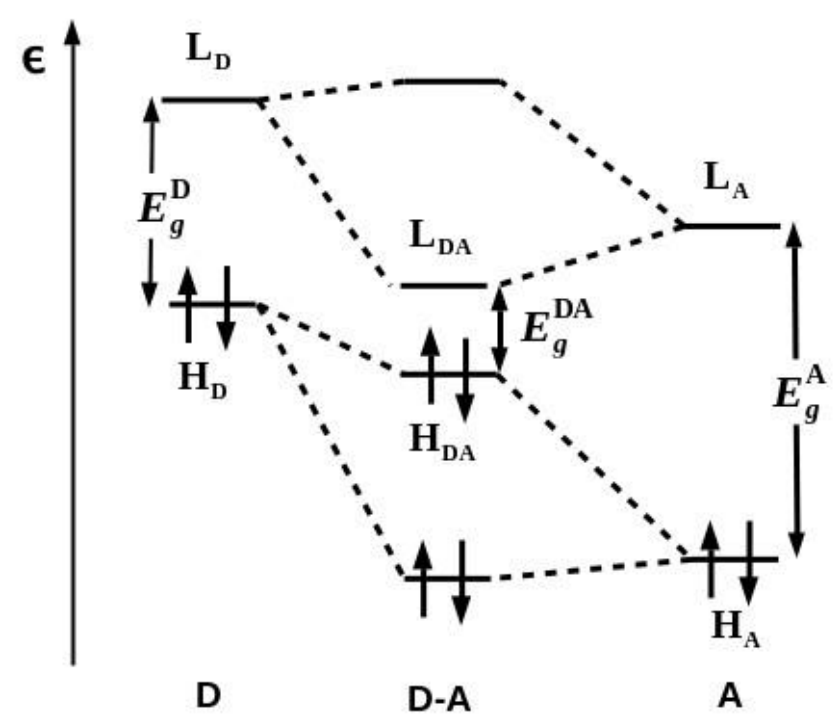

FIG. 8: Classic donor-acceptor molecular orbital diagram showing how the HOMO/LUMO gap is reduced when the donor and acceptor molecules interact.

Since we are ultimately interested in candidate molecules for the active layer of OLEDs, we must also calculate the photoluminescence (PL). This is done by assuming a fluorescence mechanism. Thus the molecule starts with its ground-state geometry and absorbs a photon putting it in a singlet state which subsequently decays to the lowest singlet state ($S_1$) before fluorescing (i.e., Kasha's rule).

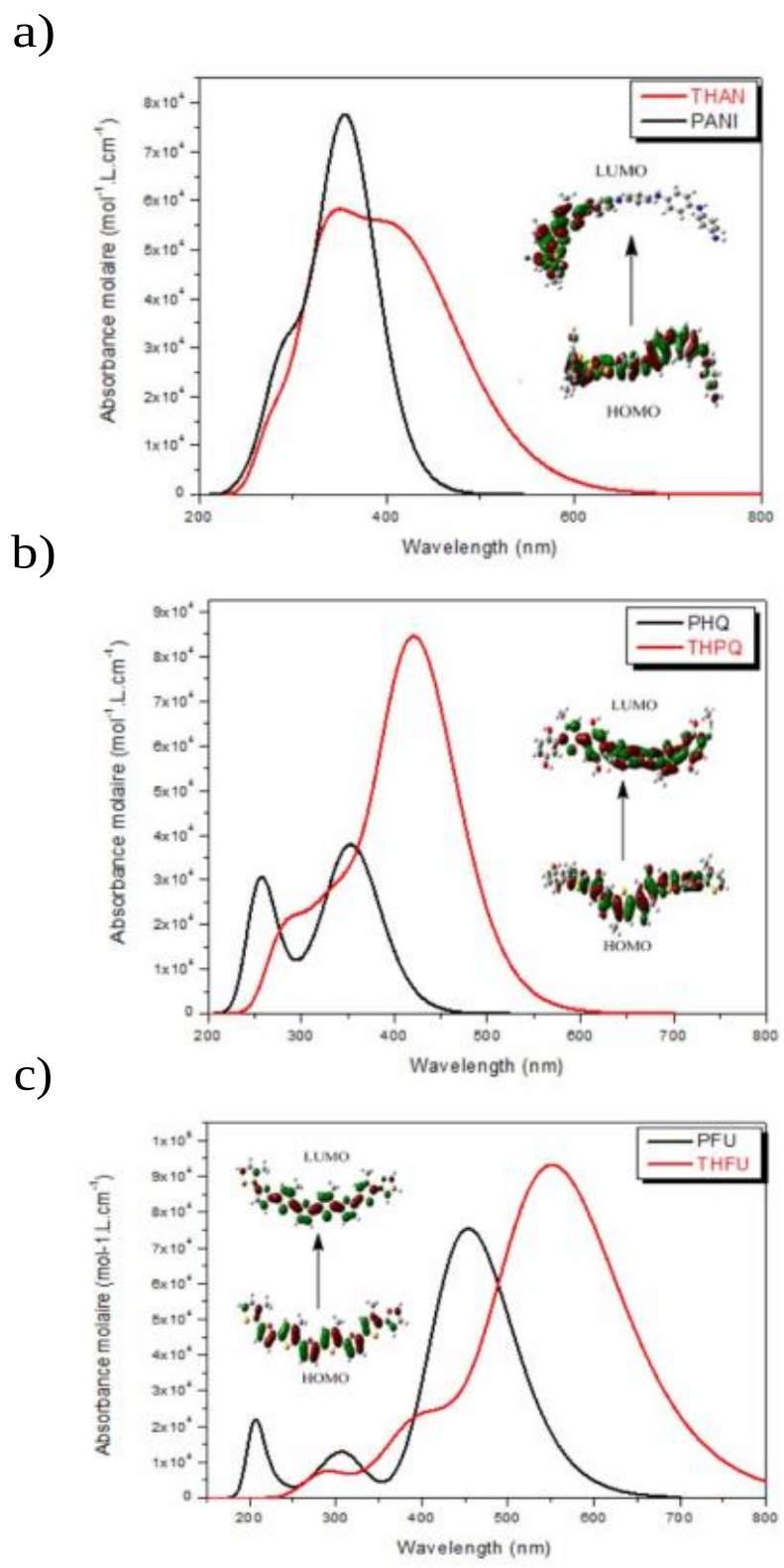

FIG. 9: TD-B3LYP/6-31G(d) UV-visible optical absorption spectra: (a) THAN and PANI, (b) PHQ and THHQ, and (c) PFu and THFu.

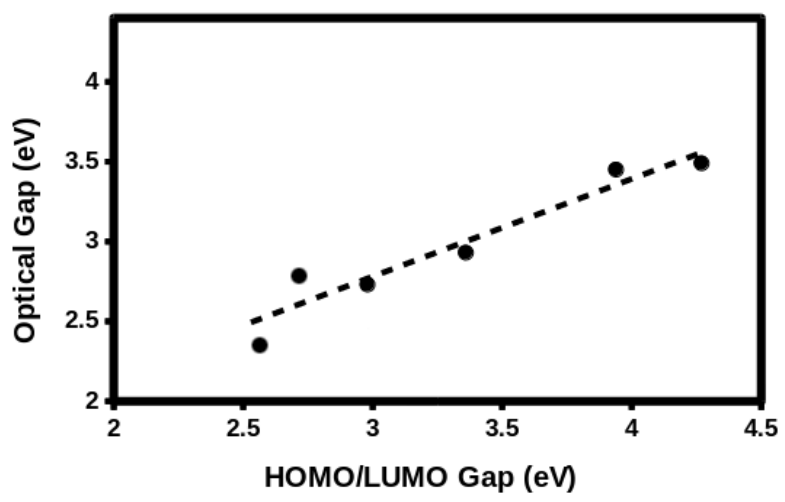

FIG. 10: Correlation between the optical gap (Table III) and the HOMO/LUMO gap (Table II).

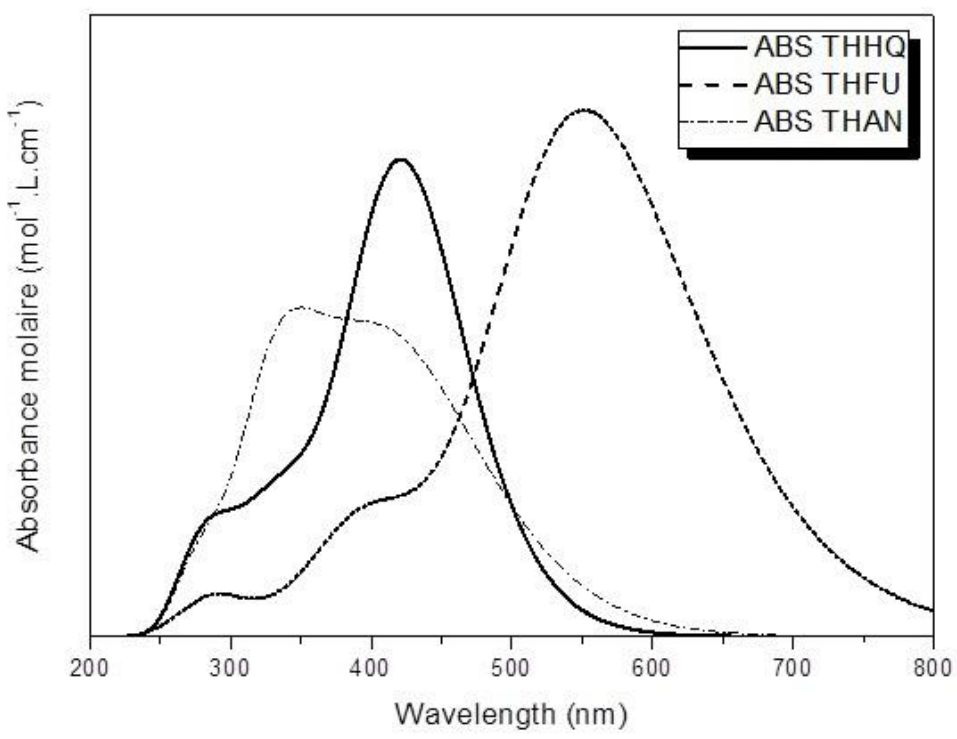

FIG. 11: TD-B3LYP/6-31G(d) UV-visible optical absorption spectra of the three co-oligomers studied in this paper.

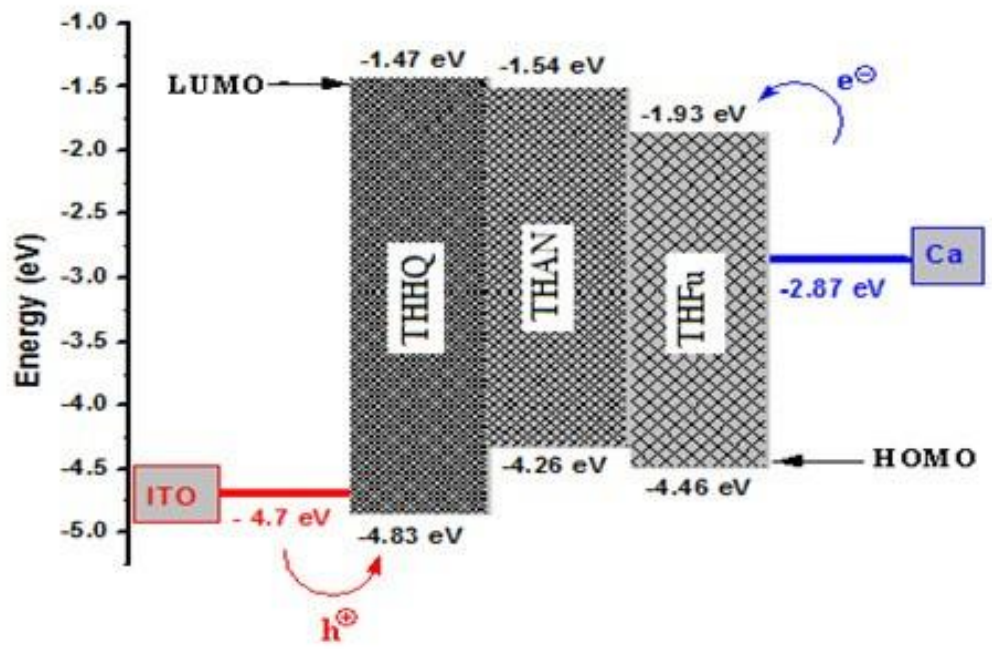

FIG. 12: Schematic energy diagram.

As the molecule has enough time for the geometry to relax in this geometry, we must recalculate the first absorption but at the relaxed $S_1$ geometry. A little reflection will show that the that the PL wavelength will always be longer than that of the absorption (the difference being the Stokes shift). Just as in the case of absorption, alternating donor and acceptor units is expected to reduce the PL energy, thus shifting it to the red. Our calculated PL spectra, shown in Fig. 3, confirm this expectation. Note that our model does not predict luminescence lifetimes nor is it accurate enough to predict the precise color of the emitted light. In general, even assuming that we have a good enough density functional, both vibrational effects and solvent effects must be taken into careful consideration when predicting colors. However shold emission take place, our calculations indicate that it should take place in or near the visible part of the spectrum.

## IV. DISCUSSION AND CONCLUSION

At the beginning of this article, we proposed to investigate the effect of alternating donor and acceptor units on emission properties. Our calculations of HOMO/LUMO gaps, optical gaps, and PL spectra all indicate that this strategy is an effective way to reduce the gap. Careful choice of the D and A units should then allow us to tailor the PL energy as we like. The present model does not allow us to predict either luminescence lifetimes or intensities, but it does allow us to say something about absorption intensities. In this case, lowering the HOMO/LUMO gap improves the intensity of the absorption spectra.

Our calculations indicate that, of the co-oligomers studied, THHQ appears to be the best candidate for an active layer of the organic electroluminescent diodes because its HOMO/LUMO gap energy is about 3.36 eV, its maximum of absorption is around 420 nm and above all, the value of its HOMO energy level is estimated at -4.83 eV (Table II), which is close to the output of the ITO anode. This facilitates the injection of the holes (Fig. 12).

## Acknowledgments

TM would like to thank the Tunisian government for grants allowing him to spend a total of 6 months in Grenoble. Grenoble Centre d'Expérimentation du Calcul Intensif en Chimie (CECIC) computing resources are gratefully acknowledged as is computing help from Pierre Girard.

## Author Contributions

This is the result of two three-month visits of Tarek Mestiri to Grenoble. The PhD thesis of Tarek Mestiri is officially directed by prof. Kamel Alimi and unofficially co-directed by prof. Mark E. Casida and by his PhD student Ala Aldin M. H. M. Darghouth. The bulk of the manuscript writing was done by Tarek Mestiri with major corrections made by Kamel Alimi and by Mark E. Casida. All authors have read and approved the final manuscript.

## Conflicts of Interest

The authors declare no conflict of interest.

## Supplementary Material

The Supplementary Material reports the calculated vibrational spectra of the molecules studied in this paper, discusses a possible synthetic route via oxidative crosslinking, and gives a summary of abbreviations used in this paper.

* Electronic address: mestiri.tarek@gmail.com
† Electronic address: ala.darghouth@univ-grenoble-alpes.fr
‡ Electronic address: mark.casida@univ-grenoble-alpes.fr
§ Electronic address: kamel.alimi@fsm.rnu.tn

# Supplementary Material for "Density-Functional Theory Study of the Optoelectronic Properties of π-Conjugated Copolymers for Organic Light-Emitting Diodes"

Tarek Mestiri*
Unité de Recherche: Matériaux Nouveaux et Dispositifs Électroniques Organiques UR 11ES55,
Faculté des Sciences de Monastir, Université de Monastir, Monastir, Tunisia

Ala Aldin M. H. M. Darghouth†
Laboratoire de Chimie Théorique, Département de Chimie Moléculaire (DCM),
Institut de Chimie Moléculaire de Grenoble (ICMG),
Université Grenoble-Alpes, 301 rue de la Chimie, CS
40700, 38058 Grenoble Cedex 9, France

Mark E. Casida‡
Laboratoire de Chimie Théorique, Département de Chimie Moléculaire (DCM),
Institut de Chimie Moléculaire de Grenoble (ICMG),
Université Grenoble-Alpes, 301 rue de la Chimie, CS
40700, 38058 Grenoble Cedex 9, France

Kamel Alimi§
Unité de Recherche: Matériaux Nouveaux et Dispositifs Électroniques Organiques UR 11ES55,
Faculté des Sciences de Monastir, Université de Monastir, Monastir, Tunisia

Contents

## I. VIBRATIONAL PROPERTIES

### A. P3HT-aniline copolymers

The main purpose of vibrational studies is to identify the different modes of Infrared vibration in order to characterize the electronic structures of the different compounds. To do this, we consider that the geometries are perfectly optimized by the DFT method. This allow us to distinguish between the vibrations in the plan and those outside the plan. The recording of the infrared spectrum reveals the different modes of vibration of the functional groups. The frequencies of the vibrations and their assignments are reported in Table I.

Infrared absorption spectra show the presence of characteristic bands of P3HT[1] and aniline[2,3] vibrations. The analysis of these spectra shows that after the combination of P3HT, some bands undergo slight changes in their positions and their intensities (Fig. 1). The main modes of aniline vibration remain in the infrared spectrum of Pani. We cite the aromatic scissoring C-H with an intensity equal to 1498 $cm^{-1}$, the stretching C-N with a high intensity of 1858 $cm^{-1}$ and the stretching N-H with a low intensity of 3595 $cm^{-1}$. In the THAN spectrum, we have observed the appearance of two bands: one attributed to the twisting C-H of P3HT with an intensity of 1237 $cm^{-1}$ and a second one attributed to the stretching of the methyl group with a low intensity of 3091 $cm^{-1}$. On the other hand, we have noticed the presence of a band with a low intensity of 1570 $cm^{-1}$, which is attributed to aromatic rocking C-H (Fig. 2).

TABLE I: Theoretical infrared assignments of PANI and THAN at the B3LYP/6-31G level.

| ν ($cm^{-1}$) PANI | ν ($cm^{-1}$) THAN | Intensity[a] | Assignment |
|---|---|---|---|
| — | 652 | m | vibration aromatic core |
| — | 823 | w | agitation C-H aromatic |
| — | 1066 | m | twisting methyl group |
| — | 1237 | s | twisting C-H P3HT |
| 1498 | — | w | scissoring C-H aromatic |
| 1570 | 1570 | w | rocking C-H aromatic |
| 1678 | 1714 | w | stretching C=C |
| 1705 | 1768 | w | scissoring N-H |
| 1813 | 1840 | s | stretching C-N |
| 1858 | 1885 | s | stretching C-N |
| — | 3091 | f | stretching methyl group |
| 3262 | 3253 | f | stretching C-H aromatic |
| 3595 | 3577 | f | stretching N-H |

[a]Abbreviations: s, strong; m, medium; w, weak.

The signal attributed to the aromatic scissoring C-H and located at 1498 $cm^{-1}$, disappears completely in the THAN composite due to the interaction of the different units. This effect is confirmed by the shift of the infrared bands, attributed to the stretching C=C (from 1678 $cm^{-1}$ at 1714 $cm^{-1}$) and the stretching C-N stretch (1813 $cm^{-1}$–1840 $cm^{-1}$) due to the presence of more conjugated chains.

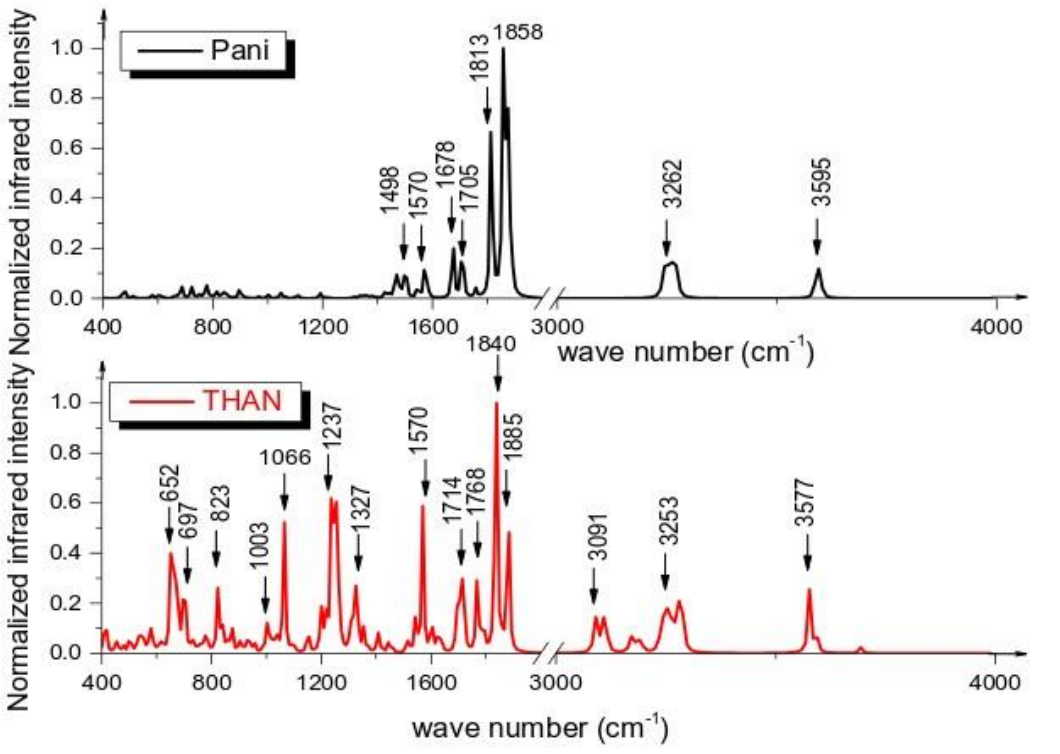

FIG. 1: Simulated infrared spectra of PANI and THAN in the ground state at the B3LYP/6-31G level.

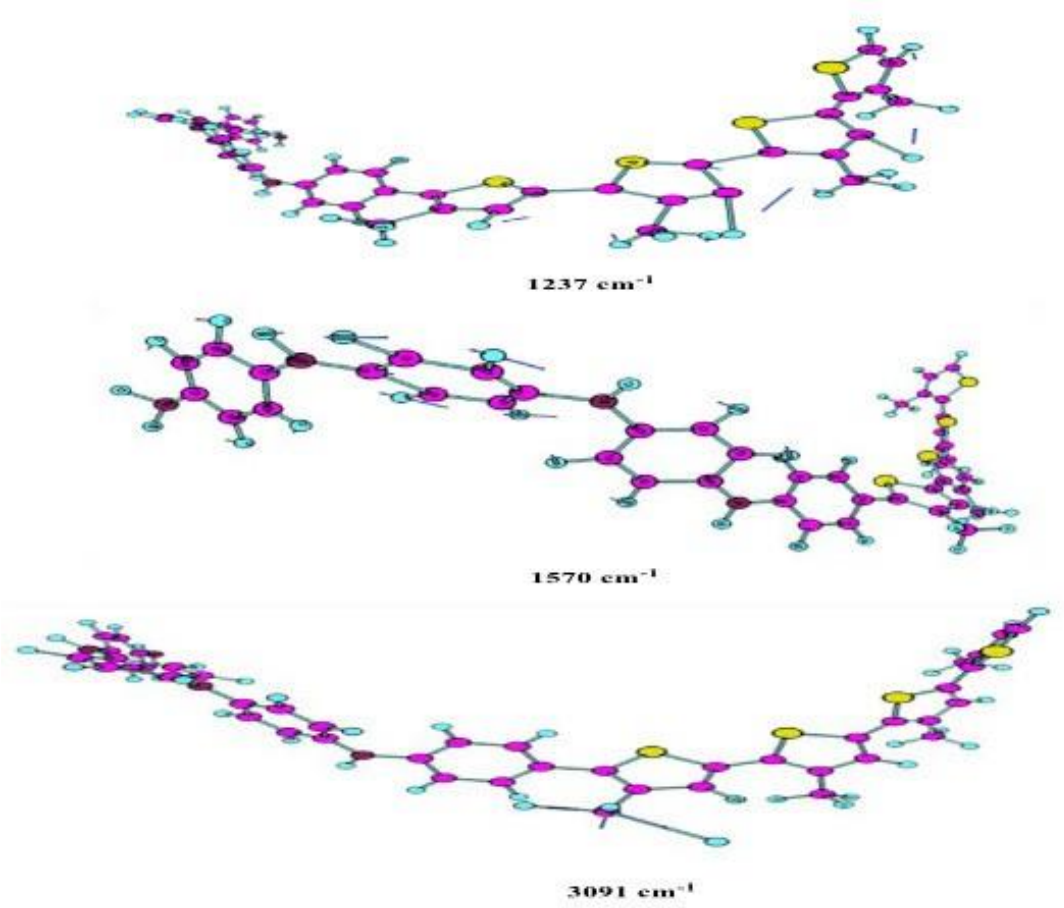

FIG. 2: Selecting the main vibrational mode of THAN.

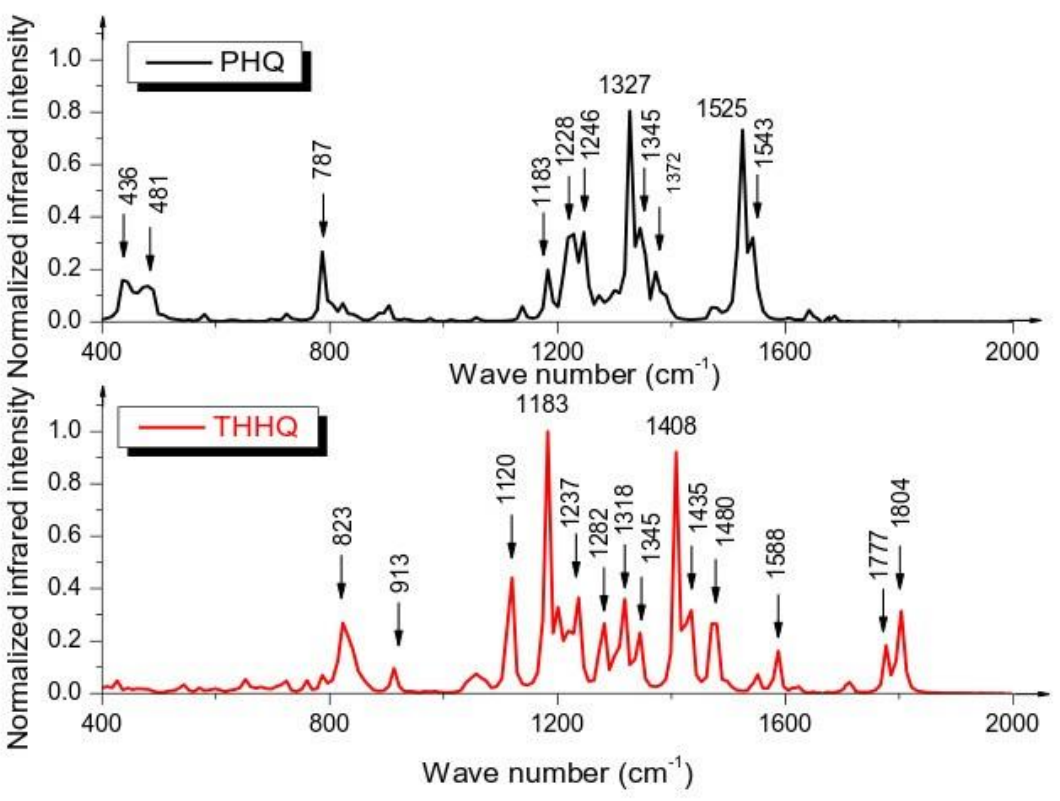

FIG. 3: Simulated Infrared spectra of PHQ and THHQ in the ground state at the B3LYP/6-31G level.

TABLE II: Theoretical infrared assignments of PHQ and THHQ at the B3LYP/6-31G level.

| ν ($cm^{-1}$) PHQ | ν ($cm^{-1}$) THPQ | Intensity[a] | Assignment |
|---|---|---|---|
| 436, 481 | — | w | twisting O-H + vibration aromatic core |
| 787 | 823 | w | stretching O-H |
| 1183 | 1183 | w | scissoring C-H aromatic |
| 1228 | 1237 | w | twisting C-H aromatic |
| — | 1282 | w | stretching C-C |
| 1327 | 1318 | s | rocking C-H aromatic |
| 1345 | 1345 | w | stretching C-C |
| — | 1408 | s | stretching C-O |
| — | 1435 | w | stretching C-C |
| 1525 | 1480 | w | stretching C=C |
| 1543 | 1588 | w | stretching C=C |
| — | 1777, 1804 | w | stretching C=C |

[a] Abbreviations: s, strong; m, medium; w, weak.

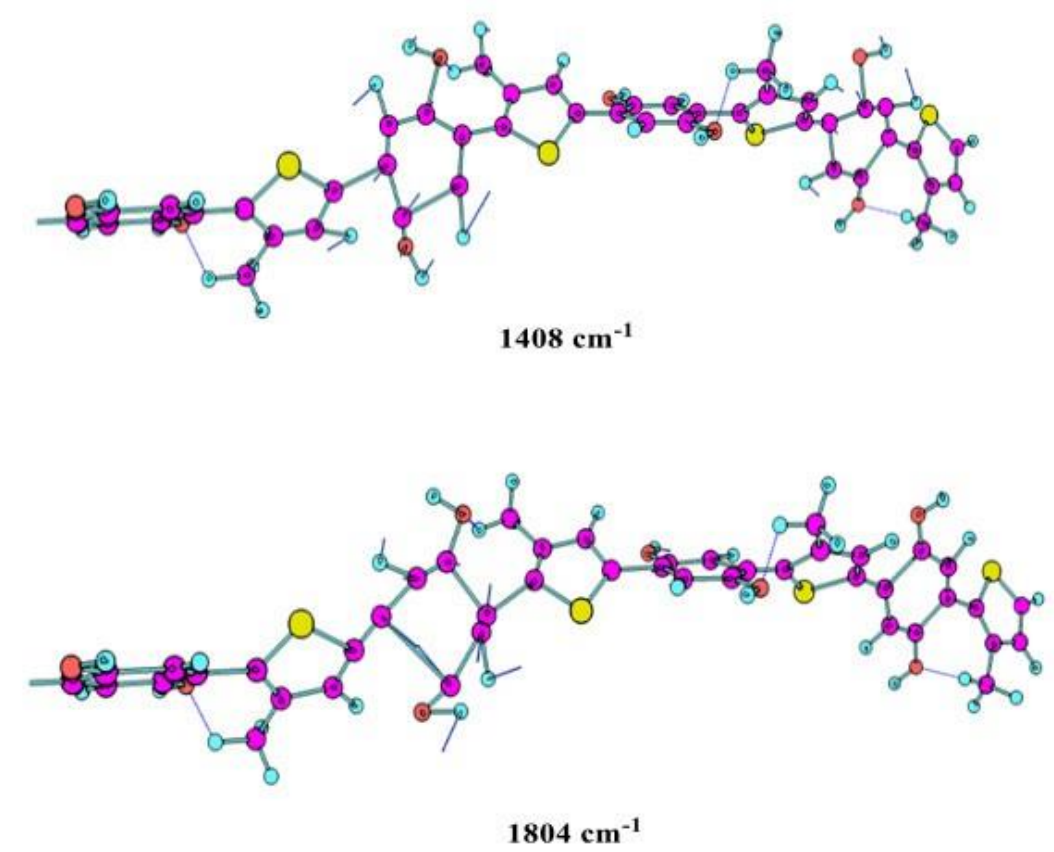

FIG. 4: Selecting the main vibrational mode of THHQ.

## B. P3HT-hydroquinone copolymers

Infrared absorption spectroscopy reveals the vibrational states of the present molecules and molecular groups. In general, the vibrational spectra depend on the structural parameters. Therefore, they can be a means to confirm the proposed structure. The IR spectrum obtained after simulation of the optimized structures of PHQ and THHQ at the ground state (Fig. 3), is presented in the following table (tab4).

The IR spectrum of PHQ is characterized by two low-intensity bands assigned to the twisting O-H with the vibration of the aromatic core at (436 $cm^{-1}$) and (481 $cm^{-1}$). There are also two more intense band s: one corresponding to the scissoring of aromatic C-H (1183 $cm^{-1}$) and another corresponding to the twisting of aromatic C-H (1228 $cm^{-1}$). Finally, we notice two bands with high intensity respectively assigned to the rocking of aromatic C-H (1327 $cm^{-1}$) and the stretching C=C (1525 $cm^{-1}$). For the THHQ spectrum, we observe a slight shift of the band in terms of position (from 1228 $cm^{-1}$ to 1237 $cm^{-1}$) attributed to the twisting of aromatic C-H. In terms of intensity, the band located at 1183 $cm^{-1}$ attributed to the scissoring of aromatic C-H becomes more intense. On the other hand, we notice the appearance of a new band with a strong intensity of about 1408 $cm^{-1}$ attributed to stretching C-O and an other less intense located at 1804 $cm^{-1}$ corresponding to the stretch ing C=C (Fig. 4). This modification is due to a significant interaction between the donor units (P3HT) and the acceptor units (hydroquinone). Thus, this comparison shows the efficient charge transfer in the polymer chain[1,4].

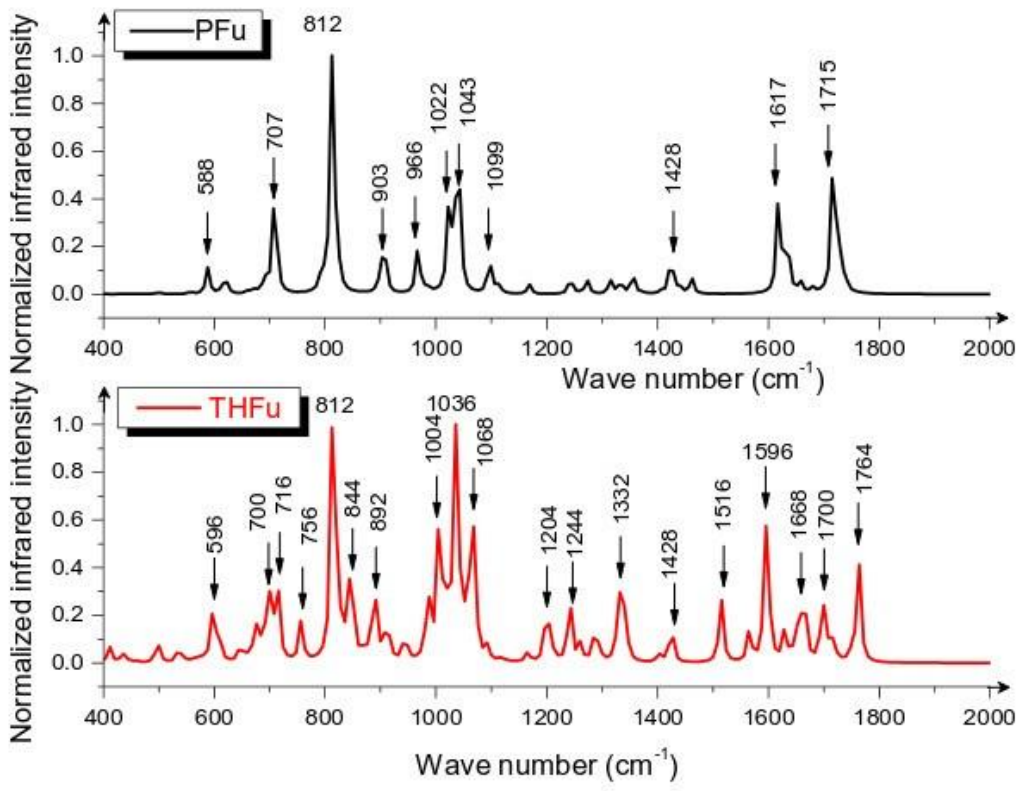

FIG. 5: Simulated Infrared spectra of PFU and THFU in the ground state at the B3LYP/6-31G level.

TABLE III: Theoretical infrared assignments of PFU and THFU at the B3LYP/6-31G level.

| ν ($cm^{-1}$) PFU | ν ($cm^{-1}$) THFU | Intensity[a] | Assignment |
|---|---|---|---|
| 588 | 596 | w | rocking C-H du furane |
| 707 | 700 | m | furane C-H vibration |
| — | 756 | w | stretching C-S |
| 812 | 812 | s | furane anti-symmetric C-H vibration |
| — | 844 | m | twisting C-H |
| 903 | 892 | w | furane core vibration |
| 966 | 948 | w | furane C-H scissoring |
| 1022 | 1004 | m | furane C-H scissoring |
| 1043 | 1036 | s | furane C-H scissoring |
| 1099 | 1068 | s | furane C-H scissoring |
| — | 1204, 1244, 1332 | w | C-H rocking + C-H stretching |
| 1428 | 1428 | w | C-C stretching |
| — | 1516 | w | methyl wag |
| 1617 | 1596 | s | C=C stretching |
| — | 1668, 1700 | w | C=C stretching |
| 1715 | 1764 | m | furane core twisting |

[a]Abbreviations: s, strong; m, medium; w, weak.

### C. P3HT-furane copolymers

The analysis of infrared spectrum shows the presence of characteristic bands of furane[5–7] and P3HT units[8] (Ref. III). For the PFu , two band s are observed in the first place: one at (707 $cm^{-1}$) relative to the agitation C-H of furane and a second one more intense located at (812 $cm^{-1}$) and relative to the anti-symmetric agitation C-H of furane. In a second place, we have noticed a low-intensity vibration of the furane core (966 $cm^{-1}$), a scissoring C-H of furane (1043 $cm^{-1}$), a stretching C=C (1617 $cm^{-1}$) and a twisting of furane core (1715 $cm^{-1}$). After the addition of P3HT units and with reference to the PFu spectrum, we notice that the bands undergo a slight change in terms of position (Fig. 5). In fact, new bands appear and they are respectively: a band relative to the rocking C-H associated with the stretching C-C (1332 $cm^{-1}$), a second one relative to the twisting of methyl group (1516 $cm^{-1}$) and a third one relative to stretching C=C (1700 $cm^{-1}$) (Fig. 6). This explains why the units of P3HT react with the units furane through a charge transfer.

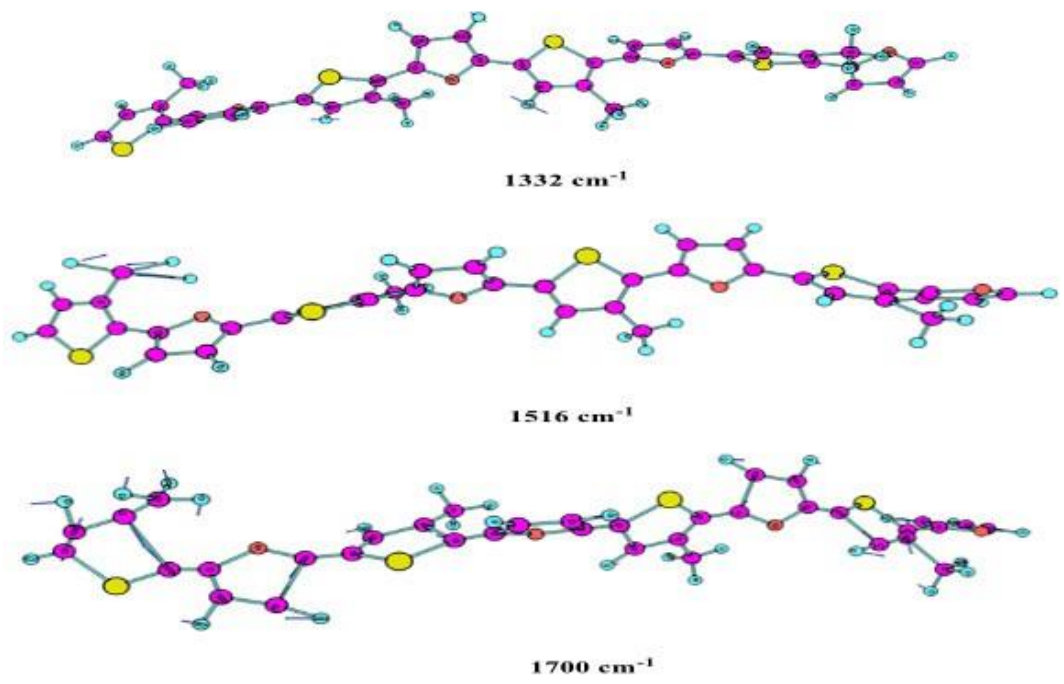

FIG. 6: Selecting the main vibrational mode of THFU.

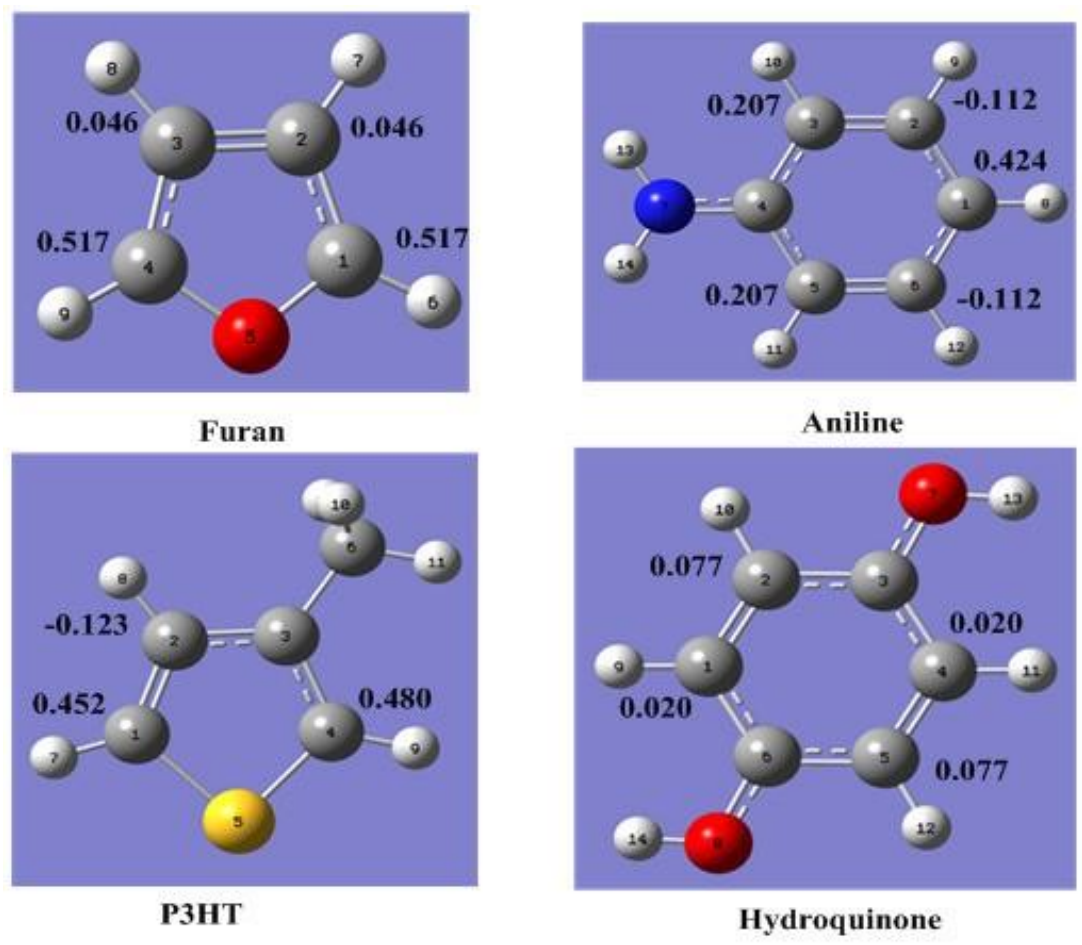

FIG. 7: B3LYP/6-31G(d,p) total atomic spin densities of radical cations of (a) furane, (b) aniline, (c) P3HT, and (d) hydroquinone units.

## II. A POSSIBLE SYNTHESIS

The Monastir group has a accumulated a good deal of experience with oxidative linking using $FeCl_3$ , so this the approach which shall be tried first. The alternating oligomers that we have in mind will be made first by electrochemical radical condensation of pairs,

$$D + A \rightarrow DA\,, \qquad (2.1)$$

followed by reaction of the pairs

$$n\ DA \rightarrow (DA)_n\,. \qquad (2.2)$$

to make oligomers and possibly polymers (depending upon the effiency of the reaction)[9–11]. Block polymers can hypothetically be made in a similar fashion by first linking

$$n\ D \rightarrow D_n$$

$$n\,A \rightarrow A_n\,, \tag{2.3}$$

and then linking the two chains together,

$$n\,D + n\,A \rightarrow (D)_n - (A)_n\,. \tag{2.4}$$

In order to get an idea of which oligomers will be easiest to make, we calculated spin densities (Fig. 7) following the same hypothesis used in previous work that the dominant reaction will occur at the sites of highest spin density. Figure 7 shows calculated spin densities. In the case of P3HT, it is noted that the highest values of the spin densities are located at the C1 and C4 carbons. These values are maximum at the carbons C1 and C7 in the case of aniline, carbons C2 and C5 in the case of hydroquinone and carbons C1 and C4 in the case of furane[12,13]. The oligomers studied have been chosen assuming linking at exactly these sites.

## III. ABBREVIATIONS

For the reader's convenience, we have collected together the abbreviations used in this paper:

A Electron accepting unit.

CB Conduction band.

D Electron donating unit.

DFT Density-functional theory.

DOS Density-of-states.

H Highest-occupied molecular orbital.

HOMO Highest-occupied molecular orbital.

ITO Indium tin oxide.

L Lowest-unoccupied molecular orbital.

LUMO Lowest-unoccupied molecular orbital.

OLED Organic light emitting diode.

P3HT Poly(3-hexylthiophene-2,5-diyl, also-known as poly(3-hexylthiophene).

PANI Oligomer of aniline.

PEDOT Polyethylenedioxythiophene.

PFU Oligomer of furane

PHQ Oligomer of hydroquinone.

PL Photoluminescence.

THAN Copolymer P3HT-aniline

THFU Copolymer P3HT-furane.

THHQ Copolymer P3HT-hydroquinone

VB Valence band.

WF Work function.

* Electronic address: mestiri.tarek@gmail.com
† Electronic address: ala.darghouth@univ-grenoble-alpes.fr
‡ Electronic address: mark.casida@univ-grenoble-alpes.fr
§ Electronic address: kamel.alimi@fsm.rnu.tn